\documentclass[preprint, amsmath,amssymb]{revtex4}
\usepackage{graphicx}
\begin{document}
\title{Frequency-tunable Superconducting Resonators via Nonlinear Kinetic Inductance}

\author{M. R. Vissers$^{1}$}
\author{ J. Hubmayr$^{1}$}
\author{M. Sandberg$^{1}$}
\altaffiliation{Present address: IBM T.J. Watson Research Center, Yorktown Heights, NY 10598, USA}

\author{S. Chaudhuri$^{2}$}
\author{C. Bockstiegel$^{3}$}
\author{J. Gao$^{1}$}
\affiliation{1) National Institute of Standards and Technology, Boulder, CO 80305 USA \\  2) Department of Physics, Stanford University, Stanford, CA 94305, USA \\ 3) Department of Physics, University of California, Sanata Barbara, CA 93106, USA }

\date{\today}

\begin{abstract}
We have designed, fabricated and tested a frequency-tunable high-Q superconducting resonator made from a niobium titanium nitride film. The frequency tunability is achieved by injecting a DC current through a current-directing circuit into the nonlinear inductor whose kinetic inductance is current-dependent. We have demonstrated continuous tuning of the resonance frequency in a 180 MHz frequency range around 4.5~GHz while maintaining the high internal quality factor $Q_i> 180,000$. This device may serve as a tunable filter and find applications in superconducting quantum computing and measurement. It also provides a useful tool to study the nonlinear response of a superconductor. In addition, it may be developed into techniques for measurement of the complex impedance of a superconductor at its transition temperature and for readout of transition-edge sensors.
\end{abstract}

\maketitle

\section{Introduction}
Superconducting microwave resonators have found important applications in astronomical detectors\cite{Day-MKID} and quantum computing\cite{Jonas-review} in the past decade. Usually these resonators are lithographically defined and have fixed resonance frequency determined by their geometry. Tunable resonators have also been developed for special applications as amplifiers\cite{Konrad-JPA,Yamamoto-Amp}, tunable couplers \cite{Martinis-tunable-coupler}, as a spectrometer for two-level systems (TLS)\cite{Micah-JJDS}, and as a probe of fundamental physics \cite{Wilson-Casimir}. Previously, tunable resonators have been typically made utilizing the nonlinear inductance of a Josephson junction as the tunable element.  This could include a long junction in a superconducting loop \cite{Osborn}, or one or more superconducting quantum interference device (SQUID) loops \cite{Laloy,Sandberg-tunares}. These devices, benefiting from the large nonlinear inductance of a Josephson junction, can achieve fractional frequency shifts of up to $50\%$, e.g., from 4~GHz to 8~GHz\cite{Konrad-JPA}. However, the critical current of the junctions, typically no more than several microamps, limits the saturation current in the resonator. These junctions also have relatively high loss, limiting the internal quality factor $\textrm (Q_i)$ of the tunable resonators to $\sim 10,000$. High $Q_i$ coplanar waveguide (CPW) resonators have been shown to be tunable through the application of a perpendicular magnetic field\cite{Healey-tunaCPW}, but the frequency shifts are limited to only several MHz, or $0.05\%$ for a 5 GHz resonator.

In this letter we present the design and experimental data of a tunable resonator based on nonlinear kinetic inductance. The device is fabricated in a single lithography step and has much greater tunability than the magnetic-field tunable CPW, while at the same time exhibits much lower loss and much higher saturation current than the Josephson junction based tunable resonators.

Superconductors have an nonlinear response to current that has been studied extensively \cite{Pippard-1950, Pippard-1953, Parmenter, Gittleman}. In general, the kinetic inductance is current-dependent, which can be expanded as \cite{Jonas-review}
\begin{equation}
L_{ki} = L_{0}[1 + (I/I_*)^2].\label{eqn:Lki}
\end{equation}
The characteristic current $I_*$, on the order of the critical current $I_c$, sets the scale of the current nonlinearity. It has been found recently that the nonlinear response in titanium nitride (TiN) and niobium titanium nitride (NbTiN) can exhibit extremely low dissipation. Resonators made from these films routinely show high $Q_i \sim 10^7$\cite{JPL-TiN}, in contrast to those seen in niobium nitride (NbN) and other superconductors\cite{Eom}. The nonlinear kinetic inductance in NbTiN and TiN provides an almost ideal Kerr medium for nonlinear superconducting devices. For example, it has been proposed for a broadband quantum-limited parametric amplifier\cite{Eom, Clint-LTD}. It is also essential for the tunable resonator device we show here.


\begin{figure}[h]
  \centering
  \includegraphics[width=\linewidth]{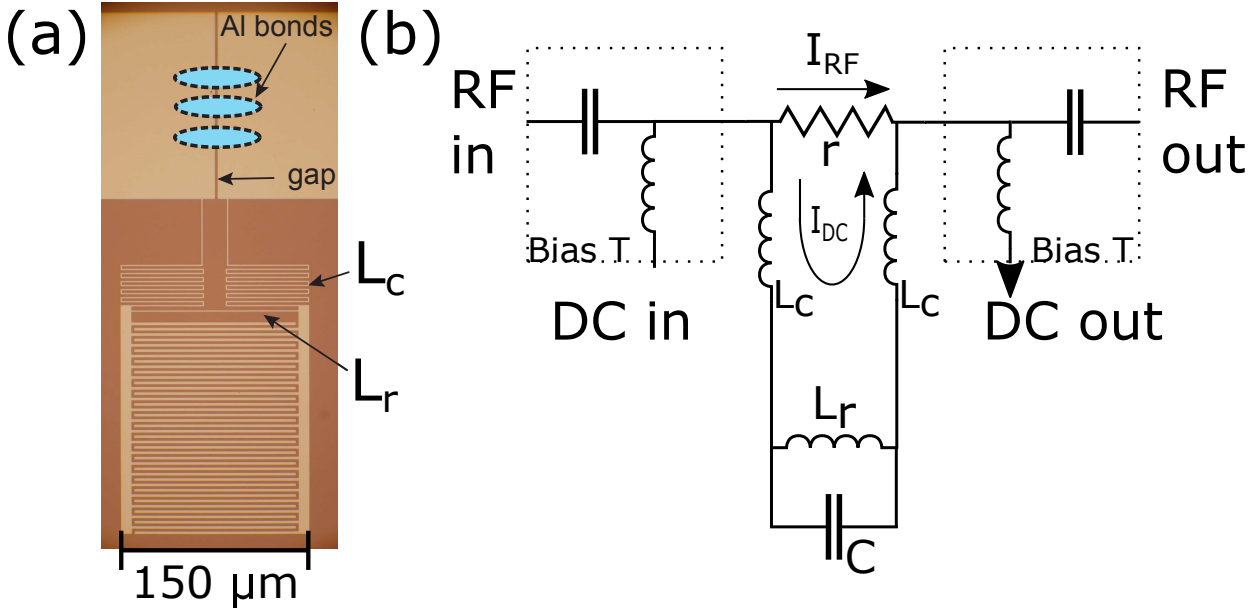}
\caption{ A micrograph of the tunable resonator device (a) and its equivalent circuit diagram (b). The LC resonance is mainly defined by a 2.5~$\mu$m wide strip inductor $L_r\approx$1~nH and a interdigitated capacitor (IDC) $C\approx$8~pF with 5~$\mu$m wide fingers and gaps. The RF choke is defined by two 2.5~$\mu$m wide meandering inductors $L_c\approx 5$~nH. A 5~$\mu$m gap in the feedline is patched by Al wirebonds as illustrated by the green ellipses in (a).}
  \label{fig1}
\end{figure}

The tunable resonator design is illustrated in Fig.~1. It is a lumped element LC resonator that is directly connected to a microstrip feedline. As can be seen from the circuit schematic Fig.~1b, the inductor $L_r\approx1$~nH and capacitor $C\approx1.2$~pF create a parallel LC resonance circuit with resonance frequency $f_r \approx \frac{1}{2\pi \sqrt{L_rC}} \approx 4.5$~GHz. The resonator is similar in design to microwave kinetic inductance detectors (MKIDs) used in astronomy \cite{BlastAPL}. For the thickness of the film and the geometry of the resonator, $L_r$ is dominated by its current-dependent kinetic inductance and serves as our tunable element. A crucial addition to this circuit that makes the resonator tunable is the gap cut in the microstrip feedline, which is then patched with normal metal (Al for $T>T_c$, or Au). If the gap did not exist or were patched with a superconductor, the device would be a standard inductively coupled resonator with fixed frequency. When a DC current $I_{dc}$ is applied to the feedline (through a pair of bias-Ts), the normal patch shows a non-zero DC resistance and since there exists a zero DC resistance path ($L_c$-$L_r$-$L_c$), the DC current is directed into the resonance frequency defining inductor $L_r$ and tunes the resonance frequency. On the other hand, a microwave tone applied to the feedline sees 1) a continuous feedline with negligible insertion loss, because the normal resistance of the patch is very small as compared to the 50 $\Omega$ characteristic impedance of the feedline, $r\ll Z_0$; 2) a very large inductive choke $L_c\gg L_r$ into the resonator, which makes the resonator weakly coupled to the feedline to achieve a desired coupling quality factor $Q_c$. Meanwhile, the large RF choke prevents the energy stored in the resonator from being dissipated in the normal patch which ensures a high resonator internal $Q_i$. When the microwave current in the resonator is small compared to the injected DC current, the fractional resonance frequency shift is given by

\begin{equation}
\frac{\delta f_r}{f_r} = -\frac{\delta L_r }{2L_r} = -\frac{I^2_{dc}}{2I^2_*}.\label{eqn:dfr}
\end{equation}

The device was fabricated from a 20~nm-thick NbTiN film. 
We co-sputtered pure Nb and Ti targets in an Ar-N$_{2}$ atmosphere at 500$^{\circ}$ C onto a high resistivity Si substrate with a 4 W RF, 100 V DC self-bias on the substrate, similar to the TiN growth reported previously\cite{Vissers-TiN-APL}. We measured $T_c\sim 13.8 K$ for the 20 nm films. 
After deposition, the devices were defined by optical photolithography using an SF$_6$ based reactive ion etch. This process was used to ensure low loss substrate-air interfaces and a high $\textrm Q_i$ resonance\cite{Sandberg-etch}.

In our experiment, the feedline patch was created by adding Al wirebonds ($T_c\sim1.2$~K) to bridge the gap in the feedline (see Fig.~1(a)). This is simple to implement as compared to adding the normal metal patch by lithography. It also gives us the opportunity to observe different effects when the feedline patch is in the superconducting ($T<T_c$) or normal ($T>T_c$) states. The device was measured at low temperatures, 60~mK - 1.4~K covering these two temperature regimes, in an adiabatic demagnetization refrigerator (ADR).
\begin{figure}[ht]
  \centering
  \includegraphics[width=\linewidth]{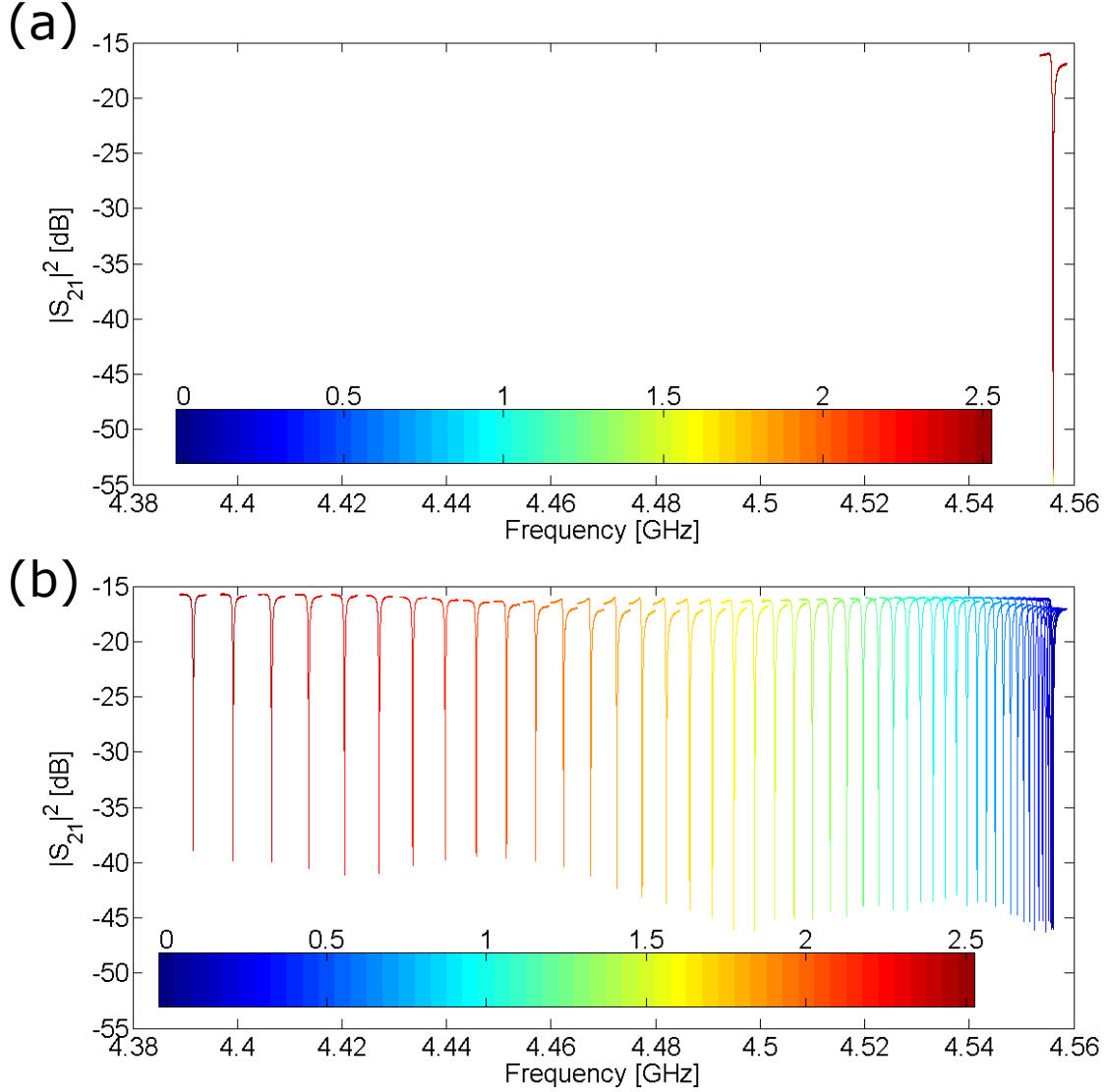}
\caption{Overlay of $S_{21}$ of tunable resontor with applied DC current 0-2.5 mA measured at 60~mK (a) and 1.4~K (b). Each color represents one $S_{21}$ trace at a separate DC current. There is no change in the frequency of the resonator with an applied DC current when the device is at 60 mK, below the $T_C$ of the Al patch, and all of the curves lie on top of each other. At 1.4~K, above the $T_C$, the frequency of the resonator as illustrated by the $S_{21}$ curves shifts down from 4.575 GHz to 4.392 GHz or 180 MHz when the DC current is applied.  }
  \label{fig2}
\end{figure}

We first measured the resonance transmission $S_{21}$ at 60~mK and 1.4~K using a vector network analyzer (VNA) while injecting a DC current to the device. The tunability of the resonator is clearly shown in Fig.~2. At 60~mK, the feedline patch is superconducting and the resonant frequency remains constant (Fig.~1a). At 1.4K, the feedline patch is normal and the resonance frequency shifts down when the DC current is increased (Fig.~2b). The moving resonance curves remain deep indicating that the resonator has maintained its high Q. With an  $I_{dc} = 2.5$~mA injected into the resonator, we have achieved a maximum frequency shift (tunability) of 180~MHz or $\delta f_r/f_r = 4\%$, which is an order of magnitude higher than achieved with the magnetic-field tuned CPW resonators. These results are in qualitative agreement with the current-directing effect of the feedline patch in the superconducting and normal states.

To further analyze the data quantatively, we applied the standard resonator fitting routine\cite{ThesisGao} to the 1.4~K $S_{21}$ data. The fractional frequency shift $\delta f_r/f_r$ and internal loss $1/Q_i$ as functions of DC current squared $I^2_{dc}$ are plotted in Fig.~3. As shown in Fig.~3a, the $\delta f_r/f_r$ data mostly follow a straight line, in agreement with the quadratic current dependence given by Eqn.~\ref{eqn:dfr}. However, we have found that a quadratic model can not fit all the data points perfectly and the fit can be much improved if a small quartic term ($I^4_{dc}$) is introduced (see the solid and dashed lines in Fig.~3a). In fact, such a quartic term is not unexpected, as the nonlinear dependence of kinetic inductance on current is more complicated than Eqn.~\ref{eqn:dfr} and a full expansion includes higher order terms (with even power index), as derived by the Parmenter theory\cite{Parmenter}. The quartic term could also be related to the excess frequency shift caused by pair breaking and quasiparticle generation under high DC current\cite{Tinkham}. We chose to fit the low current $I_{dc}<1.5$~mA data to Eqn.~\ref{eqn:dfr} (solid line in Fig.~3). From the fitting we derived $I_{*} = 10$~mA, for the 20~nm thick 2.5~um wide NbTiN strip inductor.
$I_*$ plays the same critical role as $I_c$ of a junction in a nonlinear circuit and our tunable resonator device provides a powerful tool to accurately determine this important material and geometry dependent parameter. While we also observed some increase in internal loss as the DC current is increased (Fig.~3b), this added loss is no more than $1.2\times 10^{-6}$, and the total resonator internal loss remains less than $5.5 \times 10^{-6}$ ($Q_i \gtrsim 180,000$). The increased internal loss could also be related to quasiparticle generation. From the measured $Q_i$ we can estimate the normal resistance, $r$, of the wirebonds, $r= \frac{4\omega_r (L_{c})^2}{Q_i L_r}$ where $\omega_r$ is the resonance frequency. From $\omega_r=2\pi \times 4.5~ \textrm{GHz}$ and $Q_i \gtrsim  1.8\times 10^5$, the normal resistance $r$ is estimated to be no greater than 0.04 Ohms. As similar but non-tunable NbTiN devices show higher $Q_i$, it is likely that this resistor in parallel to the resonator limits the $Q_i$. Future devices optimized with less resistive normal links could result in higher $Q_i$.

\begin{figure}[ht]
  \centering
  \includegraphics[width=\linewidth]{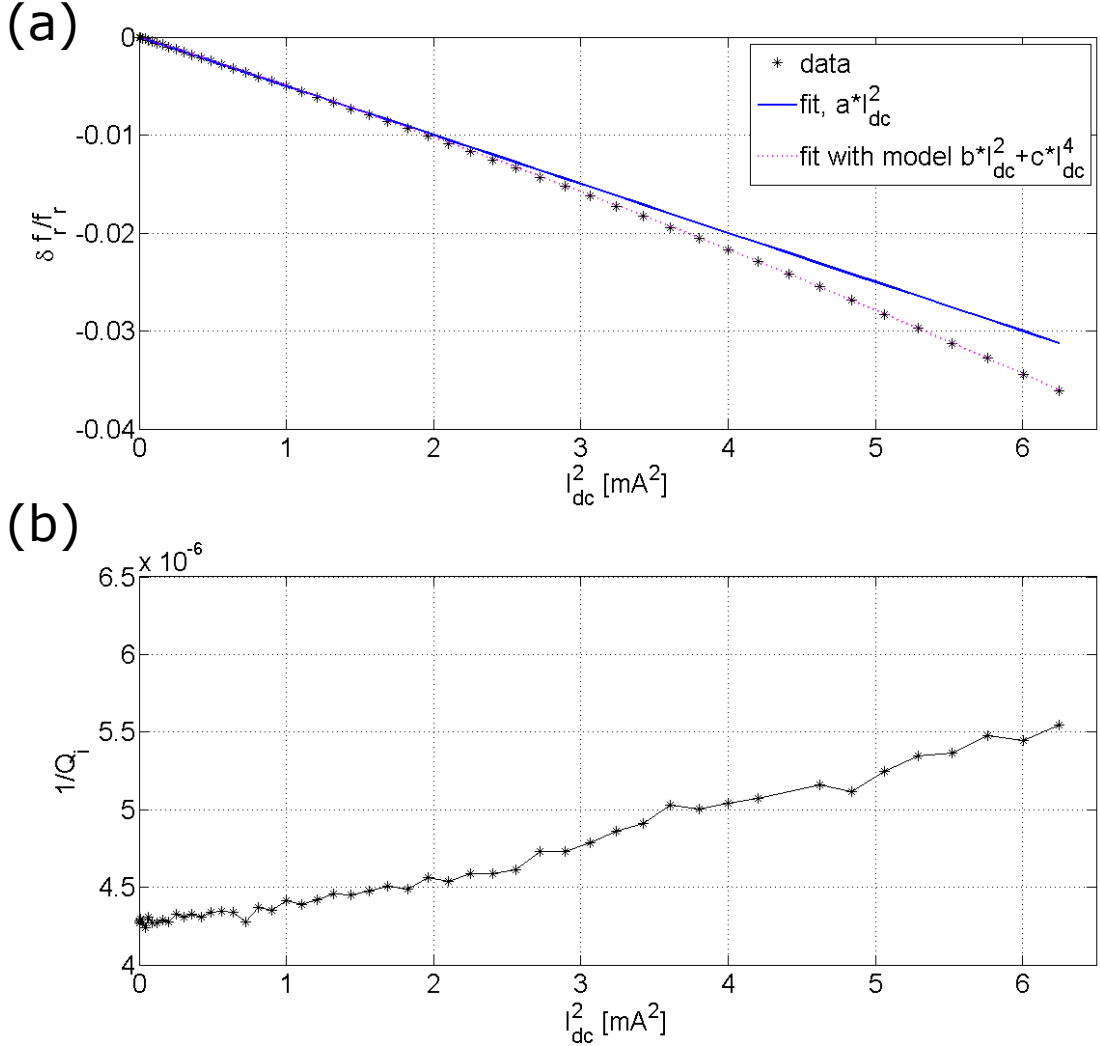}
\caption{Fractional frequency shift $\delta f_r/f_r$ (a) and internal loss $1/Q_i$ (b) vs. injected DC current into the resonator squared $I^2_{dc}$. The fitting result with and without a quartic term is indicated by the solid line and the dashed line in (a), respectively.   } \label{fig3}
\end{figure}

In the last experiment, we measured the change of $f_r$ and $Q_i$ with a constant DC current injection $I_{dc}=0.75$~mA into the resonator while the bath temperature is ramped up from 100~mK to 1.4K. The results are shown in Fig.~4. We see a steep change in both the resonance frequency and the internal loss as the temperature crosses the $T_c$ of the Al bonds ($\sim 1250$~mK), where the Al bonds are switching from superconducting state to normal state and the DC current is switching from flowing through the feedline to flowing into the resonator. With careful circuit analysis and temperature control (e.g., on-chip heater and thermometer), it should be possible to derive the complex impedance of the feedline patch to a high accuracy from the plots in Fig.~4, and therefore our tunable resonator device may be turned into a sensitive tool to measure the electrical properties of superconductor at the transition temperature. It could also be developed into a microwave readout technique for transition-edge sensors (TESs)\cite{KentTES}, if the Al wirebonds in this experiment were replaced by a TES. We are interested in exploring these directions in future experiments.

\begin{figure}[ht]
  \centering
  \includegraphics[width=8.5cm]{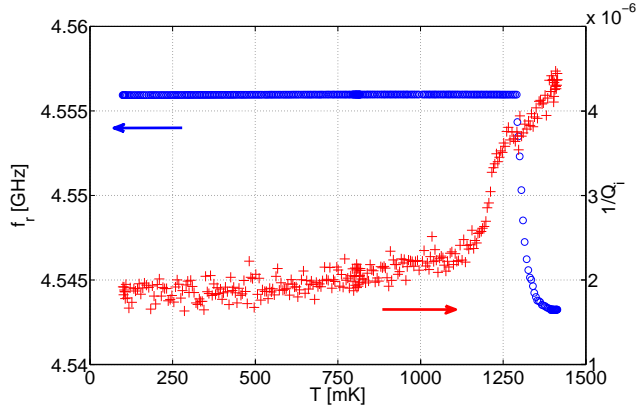}
\caption{  Frequency shift (circle) and internal loss (cross) of the tunable resonator with a constant $I_{dc} = 0.75$ mA DC current injected into the resonator as the bath temperature is increased from 100~mK to 1400~mK. Both the frequency and the internal loss see a steep change around $T_c$ of the Al patch. The amount of change is consistent with the data point corresponding to $I_{dc} = 0.75$~mA in Fig.~3.}

  \label{fig4}
\end{figure}

In conclusion we have demonstrated a high-Q resonator whose frequency can be tuned by a DC current. The resonance frequency can by tuned by $\sim 4\%$ while maintaining a high $\textrm Q_i\ge 180,000$. This device uses the current dependent nonlinearity of a high kinetic inductance superconductor NbTiN without any Josephson junctions. Compared to junction based tunable resonators, this device has much higher power handling as well as a higher $Q_i$. This device is straightforward to fabricate and can be incorporated into other superconducting circuits. The tunable resonator device has immediate applications in quantum computing and quantum measurement. For example, it can be used as notch filters to protect the qubits from the dissipative environment (Purcell filter)\cite{Purcell}, to reject the pump tone from a parametric amplifier\cite{Eom, Konrad-JPA}, and to clean the microwave tone (reduce the phase noise) applied to the microwave cavity strongly coupled to a mechanical oscillator\cite{Teufel}. Our device also provides a powerful tool to study the nonlinear response of a superconductor to current. In particular, it allows accurate extraction of $I_*$, the current scale of the nonlinear kinetic inductance, which along with the zero current term, $L_0$, leads to the full characterization of the kinetic inductance. These are important design parameters for MKIDs \cite{Loren-JAP}, parametric amplifiers, and other nonlinear devices that seek to use high kinetic inductance materials. In addition, our device may be developed to provide techniques to measure the complex impedance of a superconductor at its transition temperature and to readout transition-edge sensors. Finally, we suggest that the unique nonlinearity with extremely low dissipation of NbTiN provide an ideal Kerr medium to construct future nonlinear devices, which may find broad applications in the microwave and millimeter-wave frequency range.


%

\end{document}